\documentclass{article}
\usepackage[utf8]{inputenc}
\usepackage[a4paper, total={6.5in, 8.5in}]{geometry}
\usepackage[style=phys]{biblatex}
\usepackage{graphicx}
\usepackage{authblk}
\usepackage{hyperref}

\AtEveryBibitem{\clearfield{month}}
\AtEveryBibitem{\clearfield{doi}}
\AtEveryBibitem{\clearfield{url}}
\AtEveryBibitem{\clearfield{pages}}
\AtEveryBibitem{\clearfield{issn}}

\begin{document}

\title{Superfluid density and two-component conductivity in hole-doped cuprates}

\author[1]{Jake Ayres}
\author[2]{Mikhail I. Katsnelson}
\author[1, 2, 3]{Nigel E. Hussey}
\affil[1]{H. H. Wills Physics Laboratory, University of Bristol, Bristol BS8 1TL, United Kingdom}
\affil[2]{Institute for Molecules and Materials, Radboud University, 6525ED Nijmegen, Netherlands}
\affil[3]{HFML-FELIX, Radboud University, 6525ED Nijmegen, Netherlands}

\date{}
\maketitle

\begin{abstract}
While the pseudogap dominates the phase diagram of hole-doped cuprates, connecting the antiferromagnetic parent insulator at low doping to the strange metal at higher doping, its origin and relation to superconductivity remains unknown. In order to proceed, a complete understanding of how the single hole – initially localized in the Mott state – becomes mobile and ultimately evolves into a coherent quasiparticle at the end of the superconducting dome is required. In order to affect this development, we examine recent transport and spectroscopic studies of hole-doped cuprates across their phase diagram. In the process, we highlight a set of empirical correlations between the superfluid density and certain normal state properties of hole-doped cuprates that offer fresh insights into the emergence of metallicity within the CuO$_2$ plane and its influence on the robustness of the superconducting state. We conclude by arguing that the overall behavior is best understood in terms of two distinct current-carrying fluids, only one of which dominates the superconducting condensate and is gapped out below the pseudogap endpoint at a critical hole concentration $p^\ast$.
\end{abstract}

\section{Introduction}

At half-filling, the CuO$_4$ square-planar motif generic to all cuprates is insulating due to the strong on-site Coulomb repulsion that causes the single charge carrier within each unit cell to localize. Upon hole doping, the initial excess charge preferentially locates on the oxygen sites – giving rise to a charge-transfer insulator – which then couples antiferromagnetically with the localized charge on the copper site to create a band of Zhang-Rice singlets. With further doping, the charge-transfer gap collapses and long-range antiferromagnetic (AFM) order is destroyed, leading to a system of highly-frustrated short-range spin correlations and ultimately $d$-wave superconductivity. At the end of the superconducting (SC) dome, the charge initially localized in the Mott state becomes fully itinerant and a correlated Fermi liquid (FL) state with a large cylindrical Fermi surface containing the total carrier density emerges.

Despite a decades-long investigation, the above summary is almost the extent of our current understanding of the fate of the initially localized carrier as further holes are introduced into the CuO$_2$ plane. The starting point and endpoint are well established – it is the transition from localization to itinerancy as the number of doped holes ($p$) increases that is proving the biggest challenge. The fact that this transition incorporates the most enigmatic features of the cuprate problem; the normal state pseudogap, the strange metal and high temperature superconductivity itself makes steps to understand its evolution all the more pressing.

The aim of this Perspective is to highlight a number of intriguing correlations that shed new light, not only on the fate of the localized hole -- as reflected in both single-particle probes like angle-resolved photoemission spectroscopy (ARPES) and scanning tunneling microscopy (STM) and collective responses like transport -- but also on the impact of its enhanced mobility on the superconductivity and in particular, on the superfluid density. The totality of the transport behaviour across the cuprate phase diagram provides compelling evidence for the presence of two current-carrying fluids, one associated with coherent fermions, the other incoherent and possibly bosonic in nature. The relative occupation of these two entities is found to show a remarkably similar doping dependence to that of the superfluid density, suggesting that the latter is dominated by only one of the former. Finally, in order to motivate the development of a consistent theoretical description of this dichotomy, we highlight systems that exhibit such a two-fluid response before presenting results from a treatment of the Hubbard model on a 2 x 2 plaquette that captures a number of the essential elements seen in the cuprates. The overall picture that emerges is one that departs significantly from the standard BCS description.

\section{Experimental survey}

Despite the diverse and complex chemistry of the cuprates, the evolution of their physical (i.e. transport, thermodynamic and spectroscopic) properties across the phase diagram is found to be strikingly similar for all well-studied families. This remarkable uniformity suggests that Fermiology plays only a minor role in defining the overall behaviour, though certain unique aspects, such as the Lifshitz transition in La$_{2-x}$Sr$_x$CuO$_4$ (LSCO) or the CuO chains in YBa$_2$Cu$_3$O$_{7-\delta}$ (Y123) or YBa$_2$Cu$_4$O$_8$ (Y124), need to be fully taken into account before a complete picture can emerge. In this experimental survey, we focus predominantly on the single-layered cuprates Tl$_2$Ba$_2$CuO$_{6+\delta}$ (Tl2201) and Bi$_{2+z-y}$Pb$_y$Sr$_{2-x-z}$La$_x$CuO$_{6+\delta}$ (Bi2201) on which the bulk of our recent high-field transport studies have been performed, though for completeness, we also include complementary data obtained on LSCO, Y123, Y124 and single-layered HgBa$_2$CuO$_{6+\delta}$ (Hg1201). Moreover, in order to aid subsequent discussions, we redefine the phase diagram in Figure 1A in terms of the superfluid density $n_s(0)$, rather than $T_c$. In this way, the SC dome can be divided simply into two regimes -- pseudogapped (PG) and strange metal (SM) -- rather than the usual three (underdoped, optimally doped and overdoped).

The growth of $n_s(0)$ with increasing $p$ inside the PG regime -- known as the Uemura relation -- reflects the fact that the pseudogap itself is states-non-conserving, i.e. that states lost through the opening of the pseudogap (preferentially near the Brillouin zone edges) are not recovered below $T_c$ \cite{tallon_doping_2001}. The reduction in $n_s(0)$ beyond $p^*$, on the other hand, has commonly been attributed to pair breaking -- the so-called dirty $d$-wave scenario -- though recent studies on LSCO thin films have challenged this association \cite{bozovic_2016, mahmood_2019}. Robust correlations between $n_s(0)$ and certain key transport coefficients -- highlighted in Figure 1 and to be addressed in turn in the following sections -- provide further clues as to the origin of this loss of condensate, with the nature of the carriers within the SM regime and proximity to the Mott state at half-filling playing pivotal roles.

\subsection{Zero-field Resistivity}

In considering the in-plane resistivity $\rho_{ab}(T)$, we find it more insightful to describe the doping evolution in reverse, i.e. with decreasing $p$, starting from the far overdoped side. Beyond the SC dome ($p > p_{SC}$ where $p_{SC}$ $\sim 0.30$ is the doping at which superconductivity vanishes on the overdoped side), $\rho_{ab}(T)$ displays a pure $T^2$ resistivity \cite{manako_1992, nakamae_2003} indicative of a correlated FL ground state. As $p$ is reduced below $p_{SC}$, however, a finite $T$-linear term  emerges with a coefficient $A_1$ that grows with decreasing $p$ \cite{cooper_2009}. The striking correlation between $A_1$ and $n_s(0)$ \cite{phillips_2022} within the SM regime -- captured in panels A and B of Fig.~1  -- is reminiscent of the correlation found in conventional BCS superconductors between $T_c$ and $\lambda_{tr}$, the transport scattering rate linked to the electron-phonon coupling strength $\lambda$ and the slope of the $T$-linear resistivity (at intermediate temperatures). The fact that a similar correlation exists in both hole- and electron-doped cuprates has motivated a prolonged and intense search for a corresponding $\lambda$ in high-$T_c$ cuprates. There is an important distinction to be made here, however. In cuprates, the $T$-linear resistivity extends down to anomalously low $T$ (in marked contrast to expectations for a FL) and persists over a broad doping range (in marked contrast to expectations for a conventional quantum critical metal \cite{Hussey_2018}). Moreover, the $T$-dependence of $\rho_{ab}(T)$ in Bi2201 and LSCO was recently shown to be identical for samples with the same $T_c$ (and therefore comparable $p$) values \cite{berben_superconducting_2022}, despite the fact that the $T$- and $p$-dependencies of the spin and charge fluctuation spectra of both families are markedly different \cite{wakimoto_2007, minola_2017, peng_2018, miao_2021}. This simple result profoundly challenges claims that charge and/or spin fluctuations are responsible for the strange metallic behaviour seen in cuprates. Others have linked the ubiquitous nature of the low-$T$ $T$-linear resistivity to the vague concept of Planckian dissipation \cite{legros_universal_2019} and a gradual loss of quasiparticle (QP) integrity across the SM regime \cite{putzke_reduced_2021}, though how this QP integrity is lost remains an open question.

Plotting the temperature derivative d$\rho_{ab}$/d$T$ reveals that the evolution of $\rho_{ab}(T)$ across the SM regime is best interpreted as the sum of two distinct ($\sim T$ and $T^2$) components \cite{cooper_2009, berben_superconducting_2022}, rather than as a single component with an exponent intermediate between 1 and 2, as found, for example, in the heavy fermions.  The $T^2$ component ($A_2$) -- whose evolution across the SC dome is also captured in Fig.~1B -- is presumably a continuation of the QP-QP scattering term found beyond $p_{SC}$. Hence, a picture emerges in which the $T^2$ and $T$-linear components to $\rho_{ab}(T)$ reflect, respectively, FL and non-FL contributions to the charge transport whose relative ratio evolves smoothly across the SM regime. Whether this duality reflects the presence of additive scattering rates, real-space differentiated segments or independent conducting channels, remains to be determined, though the former is difficult to reconcile with the evolution of the in-plane magnetoresistance (MR) and low-$T$ Hall number as will be outlined in the following sections.

Acknowledging the over-simplicity of a pure two-fluid picture, one can nevertheless speculate as to which of these two fluids is the main contributor to the SC condensate. The growth in $A_1$ with decreasing $p$ must reflect either enhanced scattering from the non-FL component (the corresponding $\lambda$) or an increased occupation of the non-QP sector. The striking correlation shown in Fig.~1 between $A_1$ (a quantity determined both by scattering rate and carrier number) and $n_s(0)$ (a carrier number) suggests strongly that it is the latter. Upon entering the PG regime, $A_1$ drops (almost precipitously) while $A_2$ increases sharply. This anti-correlation of the two coefficients implies that the states responsible for the SM component are the ones that are preferentially gapped out by the opening of the pseudogap (at the zone edges). The growth in $A_2$ then reflects the increasing dominance of the QP component to the total conductivity. The overall evolution of the two coefficients indicates that upon crossing $p^\ast$, only part of the \lq strange' sector is gapped out initially. Upon further reduction in $p$, the strange fluid shrinks further and faster than the reduction of the residual \lq normal' fluid, leaving the latter once again as the dominant component (as it was for $p \sim p_{SC}$). The corresponding reduction in $n_s(0)$ with decreasing $p$ across the PG regime appears then to tie the superfluid density to that component of the normal state exhibiting signatures of non-QP transport. In order to shed more light on the nature of this non-QP sector, we turn now to consider the evolution of the in-plane MR and Hall number across the phase diagram.

\subsection{Magnetoresistance}

 The MR of most simple metals can be well described by conventional Boltzmann transport theory and follows Kohler scaling, whereby $\Delta \rho/\rho(0)$ -- the fractional change of the resistivity in a magnetic field -- is proportional to $(H/\rho(0))^2$ and where $\rho(0)$ is the corresponding zero-field resistivity. Adherence to this scaling indicates that $\Delta \rho(H,T)$ and $\rho(0,T)$ are governed by the same scattering rates. In cuprates, Kohler scaling is observed only beyond the SC dome \cite{berbenayres_2022}. Within the SM regime, however, the in-plane MR adopts an entirely different form of scaling -- so-called quadrature or $H/T$ scaling -- that has also been observed in certain iron-based superconductors close to their respective quantum critical points \cite{hayes_2016, licciardello_coexistence_2019}. Specifically, the $H^2$ dependence at low fields transforms into an unsaturating $H$-linear dependence with a $T$-independent slope. At the same time, the coefficient ($B$) of the quadratic, low-field $H^2$ MR is found to exhibit a pure 1/$T$ power-law dependence, in marked contrast to the 1/$\rho(T)$ dependence one expects from Kohler scaling. Finally, in Tl2201 and Bi2201, the MR response is found to be independent of both field-orientation and impurity scattering \cite{ayres_incoherent_2021}. While the $H^2$ to $H$-linear crossover can be reproduced within a modified Boltzmann treatment through the introduction of an impedance to cyclotron motion somewhere on the Fermi surface \cite{maksimovic_2020,grissonnanche_2021, hinlopen_2022, ataei_2022}, all other aspects of the MR response represent a radical departure from the standard semi-classical picture.
 
The fact that the MR scales with $1/T$ (and not with $1/\rho(T) = 1/(\rho_0 + \alpha_1T + \alpha_2T^2)$) within the SM regime favours the two-fluid description introduced above. While $\rho(T)$ is composed of two distinct $T$-dependent components, the MR response appears to be associated only with those carriers undergoing linear-in-$T$ scattering or dissipation. The observed correlation \cite{berbenayres_2022} between $\alpha_1$ and the magnitude of the $H$-linear MR further supports this singular association. The carriers responsible for the QP-like ($\rho_0 + \alpha_2T^2$) part of the zero-field resistivity, on the other hand, appear to play no role in the MR scaling. Of course, one cannot exclude a small orbital MR of the form predicted by Boltzmann transport in Bi2201 or Tl2201 \cite{ayres_incoherent_2021}, but this appears to be dwarfed completely by the quadrature component.
 
Across $p^\ast$, the $H$-dependence of the MR remains qualitatively unchanged. The magnitude of the $H$-linear slope continues to increase with decreasing $p$, though now its continued rise appears to be tracking the loss of carriers inside the PG regime, as reflected in the growth of $\rho_0$ and, by correspondence, $\alpha_2$ (Fig.~1C). Single-power law scaling is still observed, albeit with an exponent that changes abruptly across $p^\ast$ -- from $H/T$ to $H/T^2$ \cite{berbenayres_2022} -- mirroring the crossover to a dominant quadratic $T$-dependence in $\rho_{ab}(T)$ \cite{berben_superconducting_2022}. The sharpness of this crossover is reminiscent of the discontinuous collapse of the anti-nodal QP peak across $p^\ast$ seen by ARPES \cite{tallon_2003, chen_2019}. It should be noted that although a return to quadratic resistivity suggests a recovery of conventional FL behaviour within the PG regime, the power-law scaling and in particular the absence of the residual resistivity component $\rho_0$ in the scaling of the MR is highly anomalous. It should also be stressed that such scaling is wholly incompatible with the notion of a single fluid with multiple scattering rates.

\subsection{Hall Number}

The Hall number $n_{\rm H}$, derived from measurements of the in-plane Hall coefficient $R_{\rm H}$, serves as a convenient and informative quantifier of the mobile carrier density. For an isotropic, single-band metal $R_{\rm H} = V/n_{\rm H}e$, where $V$ is the unit cell volume and $e$ the electronic charge. In lightly-doped LSCO ($x < 0.08$) \cite{ando_evolution_2004}, $R_{\rm H}$ is approximately $T$-independent below 300 K with a value set by $n_{\rm H}=p=x$ (see Fig.~1B), indicating that initially, only the doped holes are mobile. In single particle probes, the opening of the pseudogap at $p^*$ is characterised by the formation of discontinuous Fermi arcs, preferentially located near the zone diagonals, whose arc-length also scales with $p$. Remarkably, the arcs themselves continue to host coherent quasiparticles, as evidenced by the observation of quantum oscillations \cite{proust_sebastian}). At the other end of the SC dome, $n_{\rm H}(0)=1+p$, where $n_{\rm H}(0)$ is the effective carrier number obtained from $R_{\rm H}$ in the high-field, low-$T$ limit once superconductivity has been suppressed and any $T$-dependent in-plane anisotropy \cite{abdel-jawad_anisotropic_2006} washed out \cite{putzke_reduced_2021}. The value of $n_{\rm H}(0)$ in highly overdoped Tl2201 is consistent with quantum oscillations \cite{vignolle_quantum_2008} and ARPES \cite{peets_tl_2007}. Thus, at this end of the SM regime, the hole that was originally localized in the Mott parent state becomes fully itinerant.

The crossover in $n_{\rm H}(0)$ from $p$ to 1 + $p$ was originally observed in Y123 and reported to be sharp and confined to a narrow doping region around $p^\ast$ \cite{badoux_change_2016}. In Bi2201 and Tl2201, however, this crossover was found to occur over a much broader doping range between $p^*$ and $p_{SC}$ (see Fig.~1B). While the role of the conducting CuO chains in modifying this crossover in Y123 remains to be clarified, the claim that the crossover in Tl2201 occurs beyond $p^\ast$ is strengthened by the observation that it coincides with a reduction in $n_s(0)$ with {\it increasing} not decreasing $p$ \cite{culo_possible_2021}. The fact that $n_{\rm H}(0)$ decreases upon lowering $p$ {\it before} the PG itself opens suggests a profound disconnect between single-particle and particle-particle probes within the SM regime, possibly reflecting an increasing influence of current vertex corrections, though the coincidence of the reduction in $n_{\rm H}(0)$ and the growth in $A_1$ appears to support the aforementioned picture of a gradual and preemptive loss of QP integrity with underdoping. 

The anti-correlation between $n_{\rm H}(0)$ and $n_s(0)$ across the SM regime (panels A and C of Fig.~1) is particularly striking and non-intuitive. In a clean BCS superconductor, all carriers are expected to participate in the SC condensate. As mentioned in the Introduction, the prevailing explanation for the reduction in $n_s(0)$ is pair breaking induced by disorder coupled with a rapidly diminishing pairing amplitude $\Delta$. This conventional viewpoint -- based on the Landau BCS paradigm -- fails to take into account, however, the highly anomalous nature of the normal state \cite{phillips_2022}. Moreover, the dirty $d$-wave scenario predicts a rapid reduction in $T_c$ as the normal state scattering rate increases \cite{lee-hone_2020}, in marked contrast with recent findings \cite{mahmood_2022}. Last but not least, a similar correlation, shown here in Fig.~1C for the first time, is also found between $n_{\rm H}(0)$ and $F$ -- the fraction of residual (zero-energy) states in the SC state deduced from STM \cite{tromp_2022}. The quantitative agreement between the two parameters \textit{in the same cuprate family} (Bi2201) suggests that the states contributing to $n_{\rm H}(0)$ in the field-induced normal state do not form part of the SC condensate. Conversely, those states that emerge from the SC condensate upon application of a large magnetic field do not appear to exhibit an intrinsic Hall response of their own, due to them being localized \cite{pelc_2019}, bosonic (particle-hole symmetric) or, if still fermionic, fundamentally incoherent.

\section{Theoretical considerations}

The picture emerging from this survey of recent high-field studies on hole-doped cuprates is of a transport current comprising two electron fluids -- one coherent and FL-like (albeit correlated), the other non-FL-like and of unknown origin -- whose relative ratio evolves smoothly across the phase diagram. Before discussing peculiarities in cuprates from a theoretical perspective, it is worthwhile to start with a general remark on the coexistence of two types of electronic liquids for interacting electrons in crystals. FL theory, together with its microscopic justification \cite{AGD,book_Nozieres,book_Migdal} has long been considered the \lq default' many-body theory of conducting condensed matter. Recently, the concept of a \lq non-particle' state was developed, mostly within the language of holographic duality \cite{book_Zaanen,book_Sachdev}, which is argued to be a more appropriate framework to describe cuprates and other strange metals \cite{phillips_2022}. In both cases, the normal state of the system is described as a single liquid, either as a FL or a non-FL. It is not obvious therefore whether the hypothesis of a two-liquid state is consistent with general ideas of contemporary quantum many-body theory. Being very far from a complete theoretical description of the experimental situation, we are nevertheless able to give a positive answer to this question. To this aim, we will first make a general remark and then give two examples where the coexistence of two liquids in the same many-body system is reliably established.

We proceed by discussing the emerging physics in terms of the single-particle electron Green's function \cite{AGD,book_Nozieres,book_Migdal}. To avoid possible misunderstanding, one should emphasise that this formalism contains full information on the many-particle system. In particular, it is sufficient to reproduce rigorously all equilibrium \cite{luttinger_ward,pethick} and non-equilibrium \cite{baym_kadanoff} properties of interacting fermionic systems. The simplest analysis can be done in terms of single-particle density matrix (let us stress again that, despite its \lq single-particle' name, it characterizes the whole many-particle system). One can derive the following general expression \cite{vons_kats_1989} for the accelerating action of a constant uniform electric field $\mathbf{E}$:
\begin{equation} \label{accel}
    \frac{dj_a}{dt} = e^2 E_b \sum_{\mathbf{k},n}m^{-1}_{ab}(\mathbf{k},n)\rho_n(\mathbf{k}),
\end{equation}
where $\mathbf{j}$ is the electric current, $t$ the time, $e$ the electron charge and $a,b$ are Cartesian indices,
\begin{equation} \label{mass}
m^{-1}_{ab}(\mathbf{k},n) =\hbar^{-2} \frac{\partial^2}{\partial k_a \partial k_b} \left\langle\mathbf{k}n\left|\hat{H}_0\right|\mathbf{k}n\right\rangle
\end{equation}
plays the role of inverse effective mass tensor, $\left|\mathbf{k}n\right\rangle$ and $\rho_n(\mathbf{k})$ are eigenfunctions and eigenvalues of the single-particle density matrix $\hat{\rho}$ dependent on the wave vector $\mathbf{k}$, $n$ labels different eigenstates and plays the role of a \lq band index' in the many-body description of crystalline solids \cite{vons_kats_1989}, $\hat{H}_0$ is the single-particle part of the Hamiltonian (we assume that the interaction Hamiltonian depends on coordinates only), and the $\mathbf{k}$-summation in Eq.~(\ref{accel}) is taken over the Brillouin zone. The equations (\ref{accel}), (\ref{mass}) can be considered as a real-time equivalent of the optical f-sum rule frequently used to analyse strongly correlated systems \cite{basov}.

The single-particle density matrix is nothing but the equal-time single-particle Green's function $\hat{G}$, and the latter, for energies close enough to the Fermi surface, can be decomposed into QP (that is, pole) and non-QP (incoherent) parts \cite{AGD}. These two contributions to the density matrix corresponds to two contributions to Eq.~(\ref{accel}). Rigorously speaking, the QP and non-QP parts of $\hat{\rho}$ do not necessarily commute one with one another and thus do not share a common set of eigenfunctions, but at the level of the operator $\hat{\rho}$, this decomposition is always possible and accurate.  

Half-metallic ferromagnets \cite{katsnelson2008half} provide a neat example of a physical system in which two types of electron liquids coexist naturally and rigorously. Whereas for one spin projection, the electron subsystem is metallic and FL-like, for the other, non-QP states in the gap dominate close to the Fermi surface, their density of states being non-zero above the Fermi energy only for the minority-spin gap and below it for the majority-spin gap \cite{katsnelson2008half}. Importantly, their contribution to the occupation number $\rho_n(\mathbf{k})$ is almost $\mathbf{k}$-independent and thus these states are effectively currentless \cite{vons_kats_1989}. Nevertheless, they may contribute to photoemission and scanning probe spectra, tunneling, etc. \cite{katsnelson2008half}, as confirmed experimentally \cite{HMF_exp1,HMF_exp2}. 

The case of half-metallic ferromagnets is useful as it provides us with a unique opportunity to prove the separation of electron liquids into QP and non-QP (or incoherent) components. Moreover, this separation is rigorous due to its connection to conserving quantities such as the total spin (we assume that spin-orbit effects are negligible). Unsurprisingly, the situation in cuprates is much more complicated however, and we need to find a closer analog, even if its consideration cannot be done with the same rigour. In this regard, the heavy fermion (or \lq Kondo lattice') systems \cite{Stewart,Ueda} provide a useful bridge. In these systems, the 4f or 5f electrons are mostly atomic-like and form local magnetic moments but in some regime they become itinerant, contributing to the shape of the Fermi surface, charge transport etc. In some sense, these states can be considered as \lq Abrikosov pseudofermions' -- usually a purely mathematical construction introduced as a way to represent local spin operators -- that hybridize with the true conduction-band fermions and become real \cite{Coleman_Andrei}. Therefore one needs to distinguish between large and small Fermi surface constructions with and without their contribution \cite{Ueda}. 

With this in mind, let us now return to the cuprate problem. The minimal model that is supposed to keep the essential elements of cuprate physics is the single-band $t-t'$ Hubbard model derived via mapping of density-functional computational results for various representatives of this family \cite{OKA,Pavarini}.  The corresponding Hamiltonian has the form
\begin{equation} \label{hubbard}
H=-\sum_{ij}t_{ij}c^{\dagger}_{i\sigma }c_{j\sigma }+\sum_{i}U n_{i\uparrow}n_{i\downarrow }
\end{equation}
where $i,j$ are sites belonging to the simple square lattice, $t_{ij}$ is an effective hopping that is supposed to be non-zero only for nearest- ($t$) and next-nearest-neighbours ($t'$) and $U$ is the on-site Coulomb
interaction parameter. Operators $c_{i\sigma }^{\dagger}$ ($c_{i\sigma }$) create (annihilate) fermions at site $i$ with spin $\sigma=\uparrow,\downarrow$ and $n_{i\sigma}=c_{i\sigma }^{\dagger}c_{i\sigma}$. The ratio $t'/t$ is about 0.3 for the most of cuprates, with the exception of LSCO where it is closer to 0.15. Empirically, $T_c$ grows with the value of $t'/t$ \cite{Pavarini}. The Hubbard parameter $U$ is comparable with the bandwidth, so we deal with the most theoretically demanding case of intermediately strong correlations. 

Arguably the simplest approach to correlated electrons is dynamical mean-field theory (DMFT) where the lattice problem is mapped onto a single-site effective impurity problem with self-consistent determined characteristics \cite{RevModPhys.68.13}. For cuprates, however, the minimal DMFT model should start with a 2 x 2 plaquette rather than a single site since the $d$-wave SC order parameter lives on bonds rather than on sites \cite{dwCDMFT}. There have been an enormous number of cluster DMFT calculations, the corresponding references for which can be found in Ref.~\cite{Harland20,danilov2022}. Rather than discussing numbers, we focus here on an interesting observation \cite{Harland16}; that for the values of $t'/t$ near 0.3 and a realistic value of $U$, the ground state of the minimal plaquette is degenerate, i.e. the ground states for two, three and four electrons on the plaquette coincide. As a result, the plaquette Green's function has poles at energies close to zero which may be a physical realization of the soft hidden fermion mode suggested earlier phenomenologically \cite{Imada_PRL}. This makes the system in some sense analogous to the Kondo lattice, and this degeneracy plays a role similar to spin degeneracy in the conventional Kondo problem. In particular, making the lattice from such resonant centers naturally explains the appearance of the pseudogap \cite{Harland16,Harland20}. 

In the context of the two-fluid concept discussed here, the most important theoretical observation was made in Ref.~\cite{danilov2022} at the numerically exact solution of the many-body problem for a larger, $4\times4$ size plaquette. It turns out that the binding energy of two holes per plaquette is very strongly dependent on the ratio $t'/t$ and can reach values about $0.8t \approx 0.3$ eV which is more than an order of magnitude larger than $T_c$, as shown explicitly in Fig.~\ref{fig:2}. This means that, at least, for some range of doping, holes build incoherent pairs already in the normal phase. Despite the holes being initially expected to form something similar to a charged Bose liquid, one can suppose that the interaction with the thermal bath makes this liquid rather more classical than of Bose-Einstein character \cite{wheatley}. 

Thus, in general one can propose the following phenomenological picture. The coherent or QP part of the fermionic spectrum is mostly responsible for electron transport in the normal state and completely responsible for its coherent manifestations such as quantum oscillations. The incoherent part, on the other hand, may be associated with the formation of two-hole bound states (one could call them bipolarons but we prefer to avoid any associations with phonons since they do not appear to be relevant in this simplified picture). At high enough temperatures, the Fermi liquid of coherent electron excitations acts as a dissipative environment for the \lq bipolarons' preventing their Bose-Einstein condensation. The latter can nevertheless occur as the temperature decreases leading ultimately to superconductivity, while the coherent part remains largely untouched. While this picture provides a natural explanation for the presence of the two fluids, it remains to be explored how this incoherent part can play such a prominent role in the magnetic field response \cite{ayres_incoherent_2021}. This will be the subject of future studies.

\section*{Conflict of Interest Statement}

The authors declare that the research was conducted in the absence of any commercial or financial relationships that could be construed as a potential conflict of interest.

\section*{Author Contributions}
All authors listed have made a substantial, direct, and intellectual contribution to the work and approved it for publication.

\section*{Funding}
This work was supported by the European Research Council (ERC) under the European Union’s Horizon 2020 research and innovation programme, grant agreement nos 835279-CATCH-22 (N.E.H.) and 854843-FASTCORR (M.I.K). NEH
also acknowledges the support of EPSRC (grant ref. EP/V02986X/1). Finally, J.A. also acknowledges the support of an EPSRC Doctoral Prize Fellowship (Ref. EP/T517872/1).

\section*{Acknowledgments}
We thank Milan Allan and Alexander Lichtenstein for inspiring discussions. 

\section*{Data Availability Statement}
The datasets presented in this study are tabulated in the \href{https://doi.org/10.3389/fphy.2022.1021462}{Supplementary Information}. Further inquiries can be directed to the corresponding author.

\printbibliography

\begin{figure}[h!]
\begin{center}
\includegraphics[width=6in]{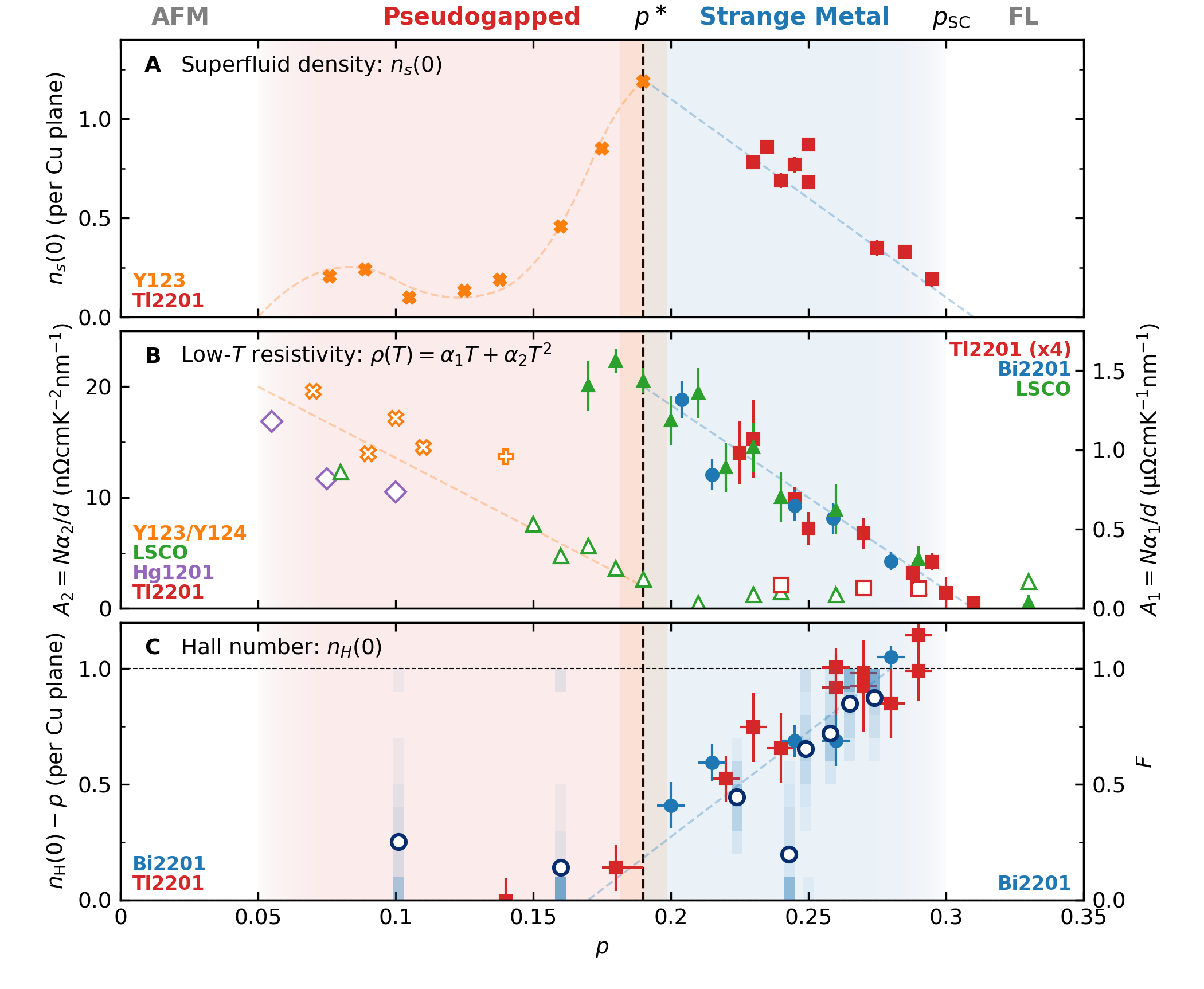}
\end{center}
\caption{Correlations between superfluid density $n_s(0)$ and various transport properties of hole-doped cuprates. \textbf{(A)} Evolution of $n_s(0)$ (per Cu) with doping $p$ in Y123 (orange crosses) and Tl2201 (red squares) obtained from measurements of the zero-temperature in-plane penetration depth $\lambda_{ab}(0)$ (summarized in Ref.~\cite{bernhard_2001, culo_possible_2021}. In order to deduce $n_s(0)$ in Y123, we used $m^\ast$ values obtained from high-field quantum oscillation studies over the doping range 0.10 $\leq p \leq$ 0.15 \cite{ramshaw_2015} and extrapolated to $p^\ast$ = 0.19 assuming that $n_s(0)$ approaches unity there. For a description of how $n_s(0)$ is obtained for Tl2201, please refer to Appendix A in Ref.~\cite{culo_possible_2021}. All parameters used in these determinations are listed in the Supplementary Material. Note that the peak in $n_s(0)$ at the pseudogap endpoint $p^\ast$ (vertical dashed line) delineates the pseudogapped (PG) regime, shaded in orange, from the strange metal (SM) regime, shaded in blue. \textbf{(B)} $p$-dependence of $A_1 = N \alpha_1 / d$ (right axis, solid points) and $A_2 = N \alpha_2 / d$ (left axis, open points), where $d$ is the $c$-axis lattice constant, $N$ is the number of CuO$_2$ planes per unit cell and $\alpha_1$, $\alpha_2$ are coefficients of the low-$T$ zero-field resistivity $\rho(T) = \alpha_1 T + \alpha_2 T^2$. Open crosses (plusses) indicate data for Y123 (Y124), respectively. See Supplementary Material for the precise values and references related to these data points. \textbf{(C)} Comparison of $n_{\rm{H}}(0) - p$, the $p$-dependence of the as-measured low-$T$ Hall carrier number of Tl2201 and Bi2201 minus its corresponding $p$ value (left axis, closed points \cite{putzke_reduced_2021}) and $F$, the fractional filling of the superconducting gap as measured recently by STM on Bi2201 (right axis, open points \cite{tromp_2022}). Shaded columns are histograms representing the local variations in the measured filling (as in Ref.~\cite{tromp_2022}). In each panel, coloured dashed lines are guides to the eye.}
\label{fig:1}
\end{figure}

\begin{figure}
\begin{center}
\includegraphics[width=0.7\linewidth]{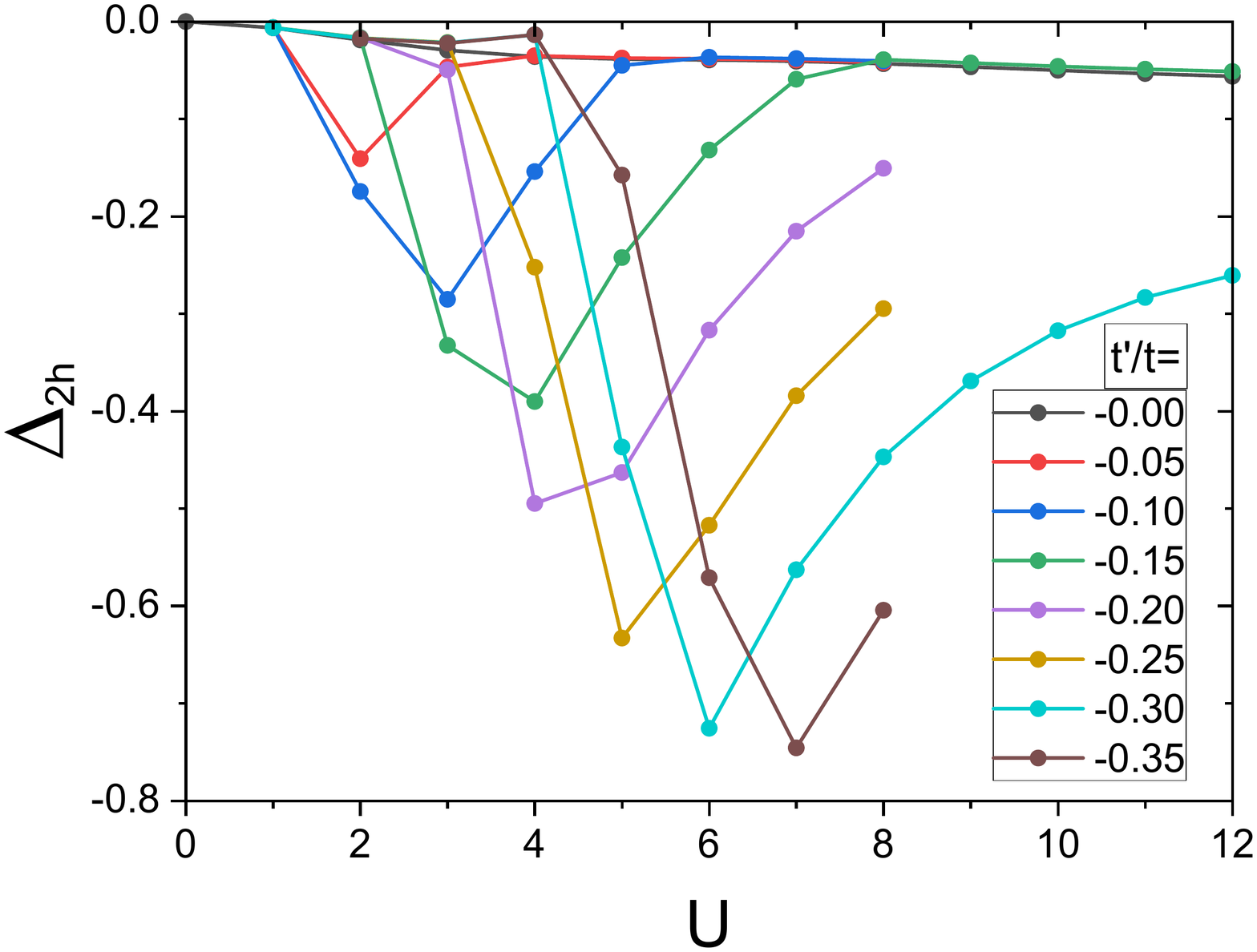}
\end{center}
\caption{Pairing energy $\Delta_{2h}$ of two holes in a 4$\times$4 cluster with periodic boundary condition as a function of $U$ and $t'$. All energies are in units of $t \approx 0.35$ eV. Reproduced with permission from Ref.~\cite{danilov2022}}
\label{fig:2}
\end{figure}

\end{document}